\documentclass[conference]{IEEEtran}

\usepackage{lipsum}
\usepackage{adjustbox}
\usepackage{url}
\usepackage[breaklinks]{hyperref}
\usepackage{breakurl}
\usepackage{mathptmx}
\usepackage{cite}
\usepackage{comment}
\usepackage{xspace}
\usepackage[usenames,dvipsnames]{xcolor}
\usepackage{tabularx}
\usepackage[ruled,vlined,linesnumbered]{algorithm2e}
\usepackage{subfig}
\usepackage{dashbox}
\usepackage{array}
\usepackage{msc}
\usepackage{flushend}
\usepackage{siunitx}
\usepackage{xspace}

\newcommand{\etal}{\textit{et al.}\xspace}

\newcommand{\ie}{\textit{i.e.,}\xspace}
\newcommand{\eg}{\textit{e.g.,}\xspace}

\renewcommand\paragraph[1]{\smallskip\textbf{#1.}}

\newcommand{\AID}{AID\xspace}

\newcommand{\ER}{Border Router\xspace}
\newcommand{\ERs}{Border Routers\xspace}
\newcommand{\er}{border router\xspace}
\newcommand{\ers}{border routers\xspace}

\newcommand{\eid}{EphID\xspace}
\newcommand{\eids}{EphIDs\xspace}
\newcommand{\id}{EphID\xspace}
\newcommand{\name}{APNA\xspace}
\newcommand{\ES}{MS\xspace}
\newcommand{\es}{ms\xspace}
\newcommand{\RS}{RS\xspace}
\newcommand{\AS}{AS\xspace}
\newcommand{\ASs}{ASes\xspace}

\newcommand{\ADlow}{accountability agent\xspace}
\newcommand{\ADslow}{accountability agents\xspace}
\newcommand{\ADcap}{Accountability Agent\xspace}

\newcommand{\ADshort}{AA\xspace}

\newenvironment{packeditemize}{
\begin{itemize}
\setlength{\topsep}{0pt}
\setlength{\itemsep}{0pt}
\setlength{\partopsep}{0pt}
\setlength{\itemindent}{0em}
}{\end{itemize}}

\begin{document}

\title{Source Accountability with Domain-brokered Privacy}

\author{
Taeho Lee, Christos Pappas, David Barrera, Pawel Szalachowski, Adrian Perrig\\
\medskip
\\
ETH Zurich\\
\medskip
\{kthlee, pappasch, david.barrera, psz, aperrig\}@inf.ethz.ch
}

\maketitle

\setcounter{page}{1}
\pagenumbering{arabic}

\begin{abstract}
In an ideal network, every packet would be attributable to its sender, while
host identities and transmitted content would remain private.  Designing such a
network is challenging because source accountability and communication privacy
are typically viewed as conflicting properties. In this paper, we propose an architecture
that guarantees source accountability and privacy-preserving communication by
enlisting ISPs as \ADslow and privacy
brokers. While ISPs can link every packet in their network to their customers,
customer identity remains unknown to the rest of the Internet. In our
architecture, network communication is based on Ephemeral Identifiers
(\eids)---cryptographic tokens that can be linked to a source only by the
source's ISP. We demonstrate that \eids can be generated and processed
efficiently, and we analyze the practical considerations for deployment.
\end{abstract}

\section{Introduction}

The commercialization of the Internet and its integral role in our daily lives
have spawned a debate on privacy and accountability---a long-standing
discussion about two properties that are typically considered conflicting.
Unfortunately, today's Internet does not provide native support for either. We
propose an architecture that resolves the accountability-privacy tussle and
guarantees network-level \textit{source accountability} and end-to-end
\textit{communication privacy}.

On one end of the spectrum, source accountability protects the integrity of the
source's identity and holds the source responsible for any traffic that it
originates. The lack of source accountability has become a Pandora's box for
Internet security. Attackers spoof their addresses and launch massive
reflection attacks exhausting the available network resources. IP source
address spoofing makes it impossible to identify the actual attacker and
renders traffic filtering ineffective, not to mention the collateral damage
when incorrectly blocking benign hosts.

On the other end of the spectrum is privacy. Recent revelations of pervasive
monitoring and mass surveillance~\cite{rfc7258} have increased user awareness
of communication privacy. Users know that their identities and network traffic
are being systematically collected by state-level entities. The lack of native
support for private communication in the Internet forces users to rely on
overlay networks and specialized applications to obtain privacy
guarantees~\cite{reed1998}. These solutions are complex to install and manage,
and degrade application performance.

To date, the research community has mainly investigated approaches that favor
either privacy or accountability, typically offering one at the expense of the
other. To our knowledge, the Accountable and Private Internet Protocol (APIP)
is the main proposal that has aimed to find a balance between the two
properties at the network layer~\cite{naylor2014}.  However, the privacy
guarantees are constrained to source anonymity; data privacy is not addressed
but delegated to conventional protocols, such as IPsec~\cite{rfc4301} and its
key exchange protocol (IKE~\cite{rfc4306}) that in themselves do not explicity
address a critical problem: certificate management (\eg issuance, revocation)
at Internet-scale.

In this paper, we propose an Accountable and Private Network Architecture
(\name) that provides strong source accountability guarantees \emph{and}
privacy-preserving communication. Our notion of communication privacy covers
host privacy (for the source and destination) and data privacy---host privacy
means that the identity (\eg IP address) of the host remains private and data
privacy means that the transmitted data remains secret from unintended
recipients.

To provide such properties, we enlist Internet Service Providers (ISPs) as a
fundamental component of our architecture for several reasons. First, we build
on past efforts to hold Autonomous Systems (ASes) accountable for malicious
traffic generated within their domain~\cite{Rowe2011, lichtman2004}. Second, we
believe ISPs have business incentives to provide privacy features to their
customers, especially in light of recent revelations regarding global
surveillance. While ISPs facilitate connection establishment between
communicating peers, encryption of traffic is still performed directly by
communication endpoints, keeping the content of the communication hidden even
from the ISPs that provides Internet connection to the peers.

In our scheme, network communication is based on Ephemeral Identifiers (\eids)
instead of long-lived network addresses, such as IP addresses. \ASs issue \eids
and assign them to their customer hosts as tokens of approval for
communication. \eids are designed to mask the host address in the network,
providing host privacy, while still providing a return address.  Preserving the
return address enables ICMP to function correctly in our scheme.  In addition,
\eids are bound to short-lived and domain-certified public/private key pairs.
These keys are used by hosts to negotiate a shared secret key, which allows
native payload secrecy through network-layer traffic encryption. 

The privacy architecture proposed in this paper, which establishes shared keys
based on \eids, by default encrypts all payload data. Pervasive encryption
frustrates large-scale traffic analysis by obfuscating all communicated
content. Moreover, payloads are encrypted with Perfect Forward Secrecy (PFS)
such that an adversary that obtains all long-term keys cannot decrypt the
content of previous communication sessions.

\eids are cryptographically linked to the identity of a host and serve as
accountability units. ISPs issue and assign \eids only to their authenticated
customers, thus bootstrapping source accountability. We argue that ISPs are the
natural \ADslow in today's Internet since they already know the identities of
their customers. Furthermore, we describe a shutoff
protocol~\cite{andersen2008}, which is a common security mechanism relying on
source accountability.  A complaining destination-host instructs an ISP to
block outgoing traffic from a customer-host that is associated with an \eid.
The accountable identifiers allow an ISP first to verify that a customer has
sent traffic to a certain destination and then to terminate any further
communication.

\paragraph{Contributions} This paper proposes a cohesive architecture,
Accountable and Private Network Architecture~(\name), that simultaneously
guarantees accountability and privacy by involving ASes as \ADslow and privacy
brokers. In particular, \name achieves the following properties:

\begin{itemize}

\item Source accountability by linking every packet in the network to its
originating source.

\item Host privacy by hiding the host's identity from every entity except the
host's AS.

\item Data privacy by supporting network-layer encryption with perfect forward
secrecy.

\item Support for feedback from the network back to the source (\eg ICMP).

\item Support for a shutoff protocol that terminates unwanted communication
sessions.

\end{itemize}

\section{Problem Definition}
Our goal is to design a network architecture that simultaneously supports
source accountability while preserving communication privacy. This section
describes the necessary requirements to realize these seemingly conflicting
goals, the security properties we strive to achieve, and the adversary models
we consider. Throughout the paper we consider that the AS of the source host
deploys \name; and in Section~\ref{ssec:name_as_a_service}, we describe how an
upstream ISP of the AS can provide \name functionalities to the host.

\subsection{Source Accountability}
\label{subsec:send_acc_def}

Source accountability refers to an unforgeable link between the identity of a
sender and the sent packet. Thus, accountability ensures that a source cannot
deny having sent a packet and a host cannot be falsely accused of having sent a
packet which it did not send. 

Achieving source accountability in practice translates to two fundamental
requirements. First, a strong notion of host identity is necessary so that
hosts cannot create multiple identities nor impersonate other hosts. Second, a
link between the source's identity and all of its traffic must be established.
This link must be established (or at least confirmed) by a third-party (\eg
source \AS) that is not the sender itself, since senders can be malicious. To
this end, the third party must observe all of the sender's traffic such that
every packet in the network can be linked to a specific sender.

\paragraph{Adversary Model} The goal of the adversary is to break source
accountability by creating a packet that is attributed to someone else in the
network. We assume that the adversary can reside in multiple ASes and that he
can see all packets within those ASes. Specifically, the adversary can
eavesdrop on all control and data messages in the network, but cannot
compromise the secret keys of the AS.

\subsection{Communication Privacy}
\label{subsec:privacy_def}

Our first goal with respect to privacy at the network layer is host privacy.
To achieve host privacy, the identity of a host must be hidden from any other
host in the source AS that is not in the same broadcast domain (\eg WiFi
network, or LAN segment) as the host,\footnote{Note that we exclude hosts in
the same broadcast domain as the host since these hosts know the \textit{Layer
2} address of the host.} any transit network that forwards traffic, as well as
the destination AS (including the communication peer). A host cannot hide from
its \AS, since the \AS knows the identity and network attachment point of every
customer. We address host privacy at the network layer, which means that
network-layer headers should not leak identity information. A host's identity
may still leak at higher layers (\eg HTTP cookies); however, these aspects are
out of scope for this paper. 

In addition, our notion of host privacy includes sender-flow
unlinkability~\cite{Pfitzmann2000}: simply by observing packet contents (both
headers and payloads) of any number of flows originating from the same \AS, the
creator(s) of the flows are no more and no less related after the observation
than they were before the observation.

Our second goal is data privacy through pervasive end-to-end encryption.
Transmitted data should be hidden from unintended recipients, including the
source and destination \ASs. To this end, the architecture must natively (\ie
without relying on upper-layer protocols) provide secure key establishment
between hosts and protection against Man-in-the-Middle (MitM) attacks.

Moreover, our notion of data privacy includes perfect forward secrecy (PFS):
disclosure of long-term secret keying material does not compromise the secrecy
of exchanged keys from past sessions and thus data privacy of prior
communication sessions is guaranteed~\cite[p. 496]{menezes2001}.

\textbf{Adversary Model:} Breaking \emph{host privacy} means that an adversary
can determine the identity of a sender, or can determine if two flows from the
same source AS originate from the same host. We assume that the adversary can
control any entity in the Internet except for the source host, hosts that are
in the same broadcast domain as the source host, and the source AS. The
adversary can observe packet headers and content, but we do not consider timing
analysis techniques, such as inter-packet arrival times.

We argue that the architecture should provide only the basic building blocks to
achieve host privacy at the network layer; and, for stronger privacy guarantees
(\eg resiliency against timing analysis), protocols at a higher layer (\eg
transport protocol) should provide such guarantees. For instance, a transport
protocol could implement a packet scheduling algorithm that homogenizes timing
between packets to prevent traffic identification algorithms based on
inter-packet timing analysis~\cite{jaber2011}. Our argument is grounded by the
fact that strong privacy guarantees often come at the expense of network
performance, and not every user (or application) requires strong privacy
guarantees. Hence, protocols that offer stronger privacy guarantees are left to
upper layers so that users can choose the appropriate protocol based on their
requirements.

An adversary can try to compromise \emph{data privacy} by decrypting the
content of a communication session between two hosts. To this end, we assume
that the adversary can control any entity in the Internet except for the two
communicating hosts and one of the two ASes that the hosts reside in.

\subsection{Additional Goals}

\paragraph{Shutoff Functionality} An accountability architecture must provide
security mechanisms that build on top of accountable addresses. A shutoff
mechanism is commonly used to terminate any active communication session
flagged for misbehavior. The architecture must ensure that the shutoff
mechanism does not create other attack vectors, such as denial of service
through non-permitted shutoff requests.

\paragraph{ICMP Support} The architecture should not sacrifice ICMP in favor of
privacy due to its importance in the Internet. It is the Swiss army knife for
network operators and is used for multiple purposes---from availability testing
(\eg ping) to network debugging (\eg traceroute) and to performance
optimizations (\eg MTU discovery).

\section{\name Overview}
This section describes the components of our Accountable and Private Network
Architecture (\name), beginning with the role of the \ASs
(Section~\ref{ssec:isps}), followed by the use of ephemeral identifiers
(Section~\ref{ssec:eids}), and ending with an end-to-end communication example
(Section~\ref{ssec:example}).

\subsection{Role of \ASs}
\label{ssec:isps}

In \name, \ASs act both as \ADslow and as privacy brokers due to their position
in the network. Since \ASs already know the identity and the physical
attachment point of their customers, they naturally act as \textit{\ADslow}. At
the same time, \ASs mask their customers' identities from all other entities,
and thus act as \textit{host-privacy brokers}.  In addition, \ASs certify their
customer-related information (\eg public keys), which is then used to generate
keys for pervasive data encryption at the network layer; thus \ASs act as
\textit{data-privacy brokers}.  We describe about each role in more detail in
the following paragraphs.

\paragraph{Accountability Functions} As an \ADlow, the \AS performs the
following functions.

First, the \AS creates a strong notion of host identity. To this end, the \AS
ensures that subscribers do not create and use multiple unauthorized identities
for their communication. \ASs already authenticate their customers and are thus
selected as \ADslow.

Second, the \AS creates a link between the identity of the source and the sent
packet.  To this end, the \AS can store every packet or insert a cryptographic
mark into every packet. Regardless of the implementation, the \AS is on the
forwarding path of all the traffic originating from its customers and is
therefore selected to establish this link. Using any other third party as an
\ADlow would require additional mechanisms to report every packet to the third
party~\cite{naylor2014}.

Third, the \AS realizes the shutoff functionality by accepting (and validating)
shutoff requests and blocking the corresponding flows. An \AS is in a strategic
position to block malicious traffic since it is close to the source and can
stop traffic before it leaves its network.

\paragraph{Privacy Functions} As a privacy broker, the \AS performs the
following functions.

First, the \AS masks the identity of its customer hosts by replacing the source
address with an ephemeral identifier (\eid). This identifier serves as a
privacy-preserving return address and thus does not break bidirectional
communication. However, \eids must be bound to specific hosts and since \ASs
know the identities of the hosts, they are well suited to perform this binding
and act as host-privacy brokers. We provide more details on \eids in
Section~\ref{ssec:eids}.

Second, the \AS acts as a certificate issuer, certifying that a public key
indeed belongs to a host in the \AS's network. More specifically, the \AS
certifies the binding between an ephemeral identifier that is issued to a host
and an ephemeral public key that is bound to the identifier. Hence, the \AS
becomes a data-privacy broker without revealing the identity of its customers.

\subsection{Ephemeral IDs}
\label{ssec:eids}

At the heart of our proposal is the use of ephemeral identifiers instead of
addresses. An \eid is an identifier associated with the identity of a host, yet
it does not leak identity information. Since \ASs know the identities of their
customers, issuing \eids to their connected hosts enables the hosts to hide
their identity without sacrificing accountability.

\paragraph{\eid as an Accountability Unit} As an accountability unit, an \eid
is an authorization token for communication that is issued by the \AS to its
customer hosts. Issuing these tokens requires strong host authentication: the
host must first prove its identity to the \AS and only then \eids can be
issued.

In \name, a host is represented to its AS through a Host Identifier (HID).  An
HID could be a hash of the host's public key or a number that is assigned by
the AS to the host (\eg IPv4 address). We do not specify how an AS assigns
HIDs, but require that HIDs be unique within the AS's boundary.  There can be
multiple \eids that are associated with an HID, and the \eids are
cryptographically bound to the HID such that only the host AS can determine the
binding.  Furthermore, an \eid serves as the accountability unit for shutoff
requests. A shutoff request against an \eid terminates all flows of the host
that use that \eid as the source identifier. In other words, flows with the
same source \eid are fate-sharing with respect to the shutoff protocol.
Blacklisting source \eids instead of source and destination \eid pairs forces
hosts to carefully manage their pool of assigned \eids.

\paragraph{\eid as a Privacy Unit} The \eid has two roles as a privacy unit: it
hides the identity of a host and provides a tool to achieve sender-flow
unlinkability. An \eid is meaningful only to the issuing AS and opaque to all
other parties. It reveals no information about the host's identity to other
hosts inside the same AS nor to the peer host that the host is communicating
with.

\eids alone are insufficient for routing packets to a destination, since
location information is missing. Therefore, a host is fully addressed by an
\AID:\eid tuple. The \AID identifies the AS in which the host resides (\eg
Autonomous System Number) and the \eid is the ephemeral identifier issued to
the host by the corresponding AS. Hence, the only leaked information is the AS
where the host resides and the host's anonymity set becomes the size of the AS
in terms of number of hosts.

In addition, decoupling the identity from the address provides a means to
achieve sender-flow unlinkability. A host can be issued multiple \eids and can
use them at will, \eg a single \eid for all flows or a different \eid for every
flow. We do not impose any requirements on how \eids are assigned. We discuss
different granularities of \eids in Section~\ref{subsec:eid_gran}.

\subsection{Communication Example}
\label{ssec:example}

\begin{figure*}[!t]
	\centering
	\includegraphics[width=.7\textwidth]{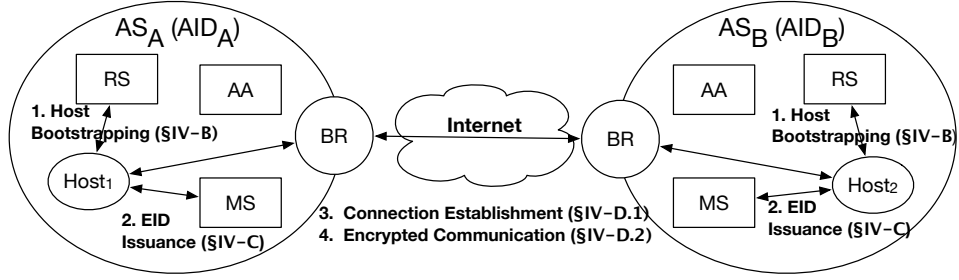}

	\caption{An end-to-end communication example.}

	\label{fig:example}
\end{figure*}

We describe the high-level workflow for communication between two hosts
(Figure~\ref{fig:example}).  The protocol details are provided in
Section~\ref{sec:details}.

The following logical entities are present in every AS: 
\begin{packeditemize}
	\item \textbf{Registry Service (RS)}: authenticates and bootstraps hosts to the AS.
	\item \textbf{Management Service (MS)}: issues \eids to the hosts.
	\item \textbf{\ER (BR)}: handles incoming and outgoing packets based on the \AID:\eid tuple.
	\item \textbf{\ADcap (\ADshort)}: handles shutoff requests against the hosts in the AS.
\end{packeditemize}

In Figure~\ref{fig:example}, a host in $\AID_A$ is trying to communicate with a host in
$\AID_B$. Communication proceeds in four steps:

\begin{enumerate}

\item \textbf{Host Bootstrapping:} the host authenticates to its \AS and
receives bootstrapping information from its \AS.

\item \textbf{\eid Issuance:} the host contacts the \ES of its AS to obtain an
\eid.

\item \textbf{Connection Establishment:} the hosts know each other's \AID:\eid
identifiers and establish a shared key that will be used for network-layer data
encryption. The shared key is derived from public keys that are associated with
the \eids.  In Section~\ref{ssec:dns_registration}, we describe how hosts can
obtain the necessary communication information through DNS.

\item \textbf{Encrypted Communication:} the hosts proceed with the actual
communication by using the corresponding \AID:\eid tuples instead of network
addresses and by encrypting every packet with their shared symmetric key.

\end{enumerate}

\section{\name Protocol Details}
\label{sec:details}
We aim to construct a lightweight architecture that avoids keeping large amount
of state on network nodes and uses symmetric cryptography for data transmission. More
specifically, we make the following design choices in \name:

\begin{enumerate}

\item symmetric encryption is used to cryptographically link \eids with HIDs;
this allows an AS to efficiently obtain the HID from the \eid without a mapping
table, which can be large;

\item proof of sending a packet is embedded in the packet, avoiding (excessive)
storage overhead for ASes; 

\item forwarding devices perform only symmetric cryptographic operations,
guaranteeing high forwarding performance.

\end{enumerate}

We begin by stating our assumptions, and proceed with the details of the
steps that are shown in our example communication scenario in
Section~\ref{ssec:example} (see Fig.~\ref{fig:example}).
Table~\ref{table:notation} summarizes the notation we use throughout the
protocol description.

\subsection{Assumptions}

\begin{itemize}

\item \textit{We assume that the cryptographic primitives we use are secure.}
For instance, we assume that the encryption scheme that protects data
communication is CCA-secure. Hence, the adversary without an encryption key
cannot learn anything about the protected plaintext from the corresponding
ciphertext, and any modification to the ciphertext by the adversary is detected
by the communicating hosts. For encrypting data communication, any conventional
CCA-secure scheme~\cite{mcgrew2004,rogaway2003} can be used. Note that
we also require that the generation of \eids to be CCA-secure, and in
Section~\ref{subsubsec:eid_struct}, we describe a CCA-secure encryption scheme
for generating \eids.

\item \textit{Participating parties can retrieve and verify the public keys of
ASes.} For example, a scheme such as RPKI~\cite{rpki} can be used to verify the
public keys of the corresponding ASes. In addition, \textit{for simplicity,
ASes use the same public/private key pairs for 1)~signing messages and 2)~key
exchanges}. In a real-world deployment, these two keys would be different, and the
key used for signing messages would be registered with RPKI.

\item \textit{Hosts do not use connection sharing devices (\eg NAT).} In other
words, each host is directly visible to its \AS. We relax this assumption in
Section~\ref{ssec:conn-sharing}.

\end{itemize}

\begin{table}[!h]
\footnotesize
\centering
\begin{tabularx}{0.45\textwidth}{rX}
    \hline\\[-5pt]
    $k_{A_i}$ & Symmetric key known among the infrastructure (\eg routers, \RS, \ES, \ADshort) within $AS_i$\\[2pt]
    $k_{H_iA_i}$ & Symmetric key shared between host $H_i$ and its AS ($AS_i$)\\[2pt]
    $k_{E_iE_j}$ & Symmetric key generated for the \eid pair $E_i$ and $E_j$\\[2pt]
	$HID_i$ & Host identifier (HID) assigned to host $H_i$\\[2pt]
    $\id_h$ & An \eid issued to host $H$\\[2pt]
    $C_{H_i}, C_{E_i}$ & Certificate for host $H_i$ and \eid $E_i$ respectively\\[2pt]
    $K_{E}^{+},K_{E}^{-}$ & Public, private key of entity $E$\\[2pt]
    $MAC_K(M)$ & Message M along with MAC of M using symmetric key $K$\\[2pt]
    $\{M\}_{K^{-}}$ & Message M along with Signature of M using private-key $K^{-}$\\[2pt]
    $E_{k}(M)$ & Symmetric encryption of $M$ with key $k$\\[2pt]
    $E^{-1}_{k}(C)$ & Symmetric decryption of $C$ with key $k$\\[2pt]
    \hline
\end{tabularx}
\caption{Notation.}
\label{table:notation}
\end{table}

\subsection{Host Bootstrapping}
\label{ssec:bootstrapping}

Initially, a host authenticates to its \AS and the bootstrapping procedure
follows thereafter. Note that host authentication is the first step towards
establishing source accountability  and is an operation that every \AS already
performs.  We do not specify how the host authenticates itself to the \AS since
well-established authentication protocols exist~\cite{rfc2865,rfc6733}. For
example, an \AS can require a user to authenticate using login credentials that
are created when the user subscribes to the \AS. During the authentication
process, we assume that the \AS learns the public key of the host ($K_H^+$).

\begin{figure}[t]
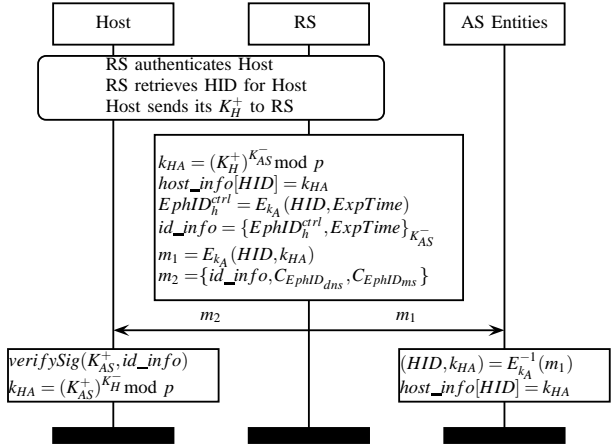

{\scriptsize
\begin{minipage}[b]{1.1\linewidth}
$\mathbf{AS\textrm{ }Entities:}$ All infrastructures (\eg \ers, \ES) of the AS\\
$\mathbf{g,p:}$ DH Parameters, $\mathbf{a:}$ A random DH Secret Integer\\
$\mathbf{generateHID():}$ Generate a unique HID\\
$\mathbf{setExpTime():}$ Create Expiration Time\\
$\mathbf{verifySig(K^+, M):}$ Verifies signature of message $M$ using $K^+$\\
\end{minipage}
\hspace*{-0.3cm}
\begin{minipage}[b]{1\linewidth}
\centering
\begin{msc}{}
\drawframe{no}
\setlength{\topheaddist}{0cm}
\setlength{\bottomfootdist}{0cm}
\setlength{\instdist}{2.6cm}
\setlength{\envinstdist}{1cm}
\declinst{host}{}{Host}
\declinst{rs}{}{\RS}
\declinst{ae}{}{AS Entities}
\nextlevel[-3.5]
\referencestart{r}{
\begin{minipage}{2.8cm}
RS authenticates Host\\
RS retrieves HID for Host\\
Host sends its $K_H^+$ to \RS
\end{minipage}
}{host}{rs}
\nextlevel[1.7]
\referenceend{r}
\nextlevel[0.4]
\setlength{\actionwidth}{4.1cm}
\setlength{\actionheight}{2.2cm}
\action{
\noindent\begin{minipage}{4cm}
    $k_{HA}=(K_H^+)^{K_{AS}^-}\textrm{mod }p$\\
    $host\_info[HID]=k_{HA}$\\
    $\id_h^{ctrl}=E_{k_A}(HID, ExpTime)$\\
    $id\_info=\{\id_h^{ctrl}, ExpTime\}_{K_{AS}^-}$\\
	$m_1=E_{k_A}(HID, k_{HA})$\\
    $m_2=$\{$id\_info, C_{\id_{dns}}, C_{\id_{\es}}$\}
\end{minipage}
}{rs}
\nextlevel[5.2]
\mess{$m_2$}{rs}{host}
\mess{$m_1$}{rs}{ae}
\nextlevel[0.5]
\setlength{\actionwidth}{2.8cm}
\setlength{\actionheight}{0.7cm}
\action{%
    \begin{minipage}{2.7cm}
    $(HID,k_{HA})=E_{k_A}^{-1}(m_1)$\\
    $host\_info[HID]=k_{HA}$
\end{minipage}
}{ae}
\setlength{\actionwidth}{2.8cm}
\setlength{\actionheight}{0.7cm}
\action{%
\begin{minipage}{2.7cm}
    $verifySig(K_{AS}^+,id\_info)$
    $k_{HA}=(K_{AS}^+)^{K_H^-}\textrm{mod }p$
\end{minipage}
}{host}
\nextlevel[1.3]
\end{msc}
\end{minipage}
}
\caption{Procedure for Host Bootstrapping.}
\label{fig:bootstrapping}
\end{figure}

Once the host has successfully authenticated, the Registry Service (\RS) of the
AS performs the bootstrapping procedure (Figure~\ref{fig:bootstrapping}).
During this procedure, the host receives information about its \AS's services
that are necessary to (later) establish communication sessions; and to support
these communication sessions, the infrastructure of the AS gets updated with
the host's information. We require that all bootstrapping messages are
authenticated in order to avoid modifications en route.

First, the \RS establishes two shared symmetric keys with the host.  One key is
used to encrypt \eid request and reply messages
(Section~\ref{ssec:ephID_issue}), and the other key is used to authenticate
every packet that the host creates and injects to the network.  The two keys
are computed by first performing a Diffie-Hellman (DH) exchange using the
public/private key pairs of the host and his AS, and then deriving the two keys
from the result of the DH exchange. Throughout the discussion, for simplicity,
we denote both keys as $k_{HA}$.

Next, the \RS creates a control \eid ($\id^{ctrl}_h$) for the host. The host
uses the control \eid to access the AS's services. For instance, using his
$\id^{ctrl}_h$, the host accesses the \ES to request data-plane \eids. Both
control and data-plane \eids are constructed identically
(See~\ref{ssec:ephID_issue}), but they are used differently and have different
expiration times. A control \eid is used for communication with the AS's
internal services and has longer lifetime (\eg DHCP lease time) while a
data-plane \eid is mainly used for data communication and is valid for the
duration of a communication session. Using the same construction for both \eid
types simplifies communication in \name: all communication is based on \eids.
For the paper, we use the term \eids to refer to the data-plane \eids.

The \RS returns the following information to the host: the control \eid
($\id^{ctrl}_h$) with its expiration time ($ExpTime$), and the certificates for
the \ES ($\id_{\es})$ and DNS ($\id_{dns}$) services. These certificates
contain \eids, which are used as the destination identifiers to access the
corresponding services, the expiration times for the \eids, and the public keys
that are associated with the respective \eids.

Finally, the \RS sends the host information ($HID$, $k_{HA}$) to infrastructure
entities in the AS (\eg routers, \ES, \ADshort); the entities store the
information in their database ($host\_info$). The infrastructure of the AS must
learn the host information in order to handle packets that are originating from
and destined to this host.  Specifically, the entities need to learn the HID of
the host ($HID$) and the shared key ($k_{HA}$) with the host so that they can
verify the authenticity of the packets that originate from the host.

\subsection{Ephemeral ID Issuance} 
\label{ssec:ephID_issue} 

An \eid is an encrypted token using the AS's secret key ($k_A$); it contains
the host's $HID$ and an expiration time that indicates the validity period for
the \eid (Equation~\ref{eq:ephID}). Note that the use of encryption enables the
issuing AS to obtain the HID and expiration time from an \eid in a stateless
fashion, without an additional mapping table; this allows an AS to handle an
arbitrary number of \eids at a constant cost.

\begin{equation}
\eid = E_{k_{A}}(HID,~ExpTime)
\label{eq:ephID}
\end{equation}

Every \eid is associated with a public/private key pair ($K_{\id}^{+},
K_{\id}^{-}$), which serves two purposes: 1)~to create a shared key with a peer
host for data encryption (Section~\ref{ssec:conn_est}), and 2)~to authenticate
shutoff requests (Section~\ref{ssec:shut_off_protocol}). Since the key pair is
used by the host to create a data encryption key that is kept secret from the
AS, it is generated by the host. 

The AS certifies the binding between an \eid and a public/private key pair by
issuing a short-lived certificate ($C_{\id}$) that has the same expiration time
as the \eid. From the certificate, a peer host learns the public key
($K_{\id}^+$) that is associated to the \eid as well as the expiration time for
the \eid. In addition to the information about the \eid, the certificate
contains information about the issuing AS---the \AID and the \eid of the \ADlow
($\id_{aa}$). The agent's \eid is used by a peer host (with which the
requesting host communicates) to initiate the shutoff protocol when necessary.

To obtain an \eid, the host creates and sends an \eid request message to the
\ES. Specifically, the host first generates the public/private key pair
($K_{\id}^{+}, K_{\id}^{-}$) for the \eid and includes $K_{\id}^{+}$ in the
request message. In addition, the host uses $\id_h^{ctrl}$ as the source
address for the request message and encrypts the message using the shared key
with the AS ($k_{HA}$). The message is encrypted to hide it from other entities
in the AS that are not part of the AS infrastructure. If an adversary who
tries to compromise sender-flow unlinkability (for the description of the
adversary model, see Section~\ref{subsec:privacy_def}) can see the content of
\eid request packets, he can identify a common sender across multiple flows at
the level of $\id_h^{ctrl}$ as the initial packets to establish connections
between two hosts contain $K_{\id}^{+}$ information for key negotiation
purposes (Section~\ref{ssec:conn_est}). That is, the adversary learns the
($\eid_{h}^{ctrl}$, $K_{\id}^{+}$) pair from \eid request packets and searches
for connection establishment packets that contain $K_{\id}^{+}$s that maps to
the same $\eid_{h}^{ctrl}$. Note that the adversary has not compromised the
host identity since only the host's AS can extract host identity from
$\eid_{h}^{ctrl}$. Nonetheless, he has successfully identified a common sender
across multiple flows. In \name, encrypting the \eid request message prevents
such attacks.

Upon receiving the request, the \ES validates the authenticity of the request;
decrypts the source \eid ($\id_h^{ctrl}$); and performs the following checks:
1)~$\id_h^{ctrl}$ has not expired, 2)~the client's identifier ($HID$) is valid
(\ie has not been revoked), and 3)~the message is valid (\ie the message can be
decrypted successfully). If any one of the checks fails, the request is
dropped.

Then, the \ES proceeds with the \eid issuance: it generates an \eid and creates
the short-lived certificate ($C_{\id}$) for the \eid. Finally, the \ES encrypts
the certificate and sends it to the requesting host. The certificate is
encrypted so that an adversary cannot relate different \eids to the control
\eid of the requesting host by observing the content of the \eid reply
packets.

\begin{figure}[t]
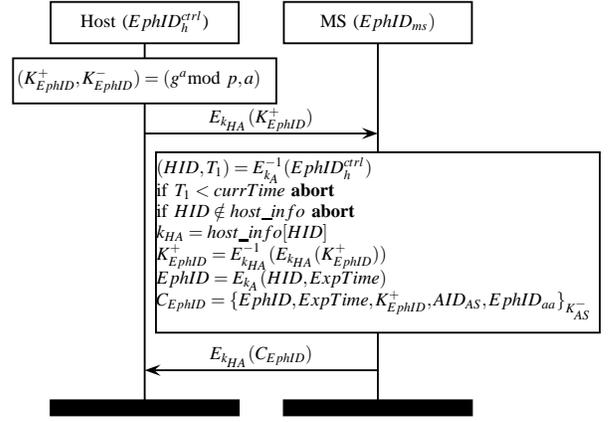

{\scriptsize
\hspace*{-0.6cm}
\begin{minipage}[b]{1\linewidth}
\raggedright
\begin{msc}{}
\drawframe{no}
\setlength{\topheaddist}{0cm}
\setlength{\bottomfootdist}{0cm}
\setlength{\instdist}{3.1cm}
\setlength{\envinstdist}{2.8cm}
\setlength{\instwidth}{2.5cm}
\declinst{host}{}{Host ($\id^{ctrl}_h$)}
\declinst{es}{}{\ES ($\id_{\es})$}
\nextlevel[-3.4]
\setlength{\actionwidth}{3.5cm}
\action{
\begin{minipage}{3.4cm}
    $(K_{\id}^{+}, K_{\id}^{-})=(g^a\textrm{mod }p, a)$
\end{minipage}
}{host}
\nextlevel[2]
\mess{$E_{k_{HA}}(K_{\id}^{+})$}{host}{es}
\nextlevel[0.5]
\setlength{\actionwidth}{5.9cm}
\setlength{\actionheight}{2.4cm}
\action{%
\begin{minipage}{5.8cm}
    $(HID, T_1)=E^{-1}_{k_{A}}(\id^{ctrl}_{h})$\\
    if $T_1 < currTime$ \textbf{abort}\\
    if $HID \notin host\_info$ \textbf{abort}\\
    $k_{HA}=host\_info[HID]$\\
    $K^{+}_{\id} = E^{-1}_{k_{HA}}(E_{k_{HA}}(K^{+}_{EphID}))$\\
    $\id=E_{k_{A}}(HID, ExpTime)$\\
    $C_{\id}=\{\id,ExpTime,K_{\id}^{+},\AID_{AS},\id_{aa}\}_{K_{AS}^{-}}$\\
\end{minipage}
}{es}
\nextlevel[5.8]
\mess{$E_{k_{HA}}(C_{\id})$}{es}{host}
\end{msc}
\end{minipage}
}
\caption{Procedure for \eid Issuance.}
\label{fig:id-issue}
\end{figure}

\subsection{Data Communication}
\label{ssec:data_comm}

To communicate, two hosts first generate a shared symmetric key for their
communication session. This key is then used to encrypt all traffic that
belongs to this communication session. We emphasize that two hosts can create
multiple communication sessions and each session has a different symmetric key
to ensure that disclosure of one encryption key does not compromise data
privacy of other communication sessions. We provide further details.

\subsubsection{Connection Establishment}
\label{ssec:conn_est}

For every connection establishment between a pair of hosts, the two hosts
perform the following tasks: 1)~verify each other's \eid certificate that is
issued by their corresponding ASes, and 2)~establish a shared key via a DH key
exchange to encrypt their communication.

Consider two hosts, \textit{A} and \textit{B}, with \eids $\id_a$ and $\id_b$,
respectively, that are trying to establish a connection with each other.
Assume that the hosts have obtained each other's \eid and the associated
certificate (we discuss obtaining \eids through DNS in Section~\ref{ssec:dns_registration}).  Using the
short-lived certificate of $\id_b$ and the public-private key pair associated
with $\id_a$, \textit{A} derives a shared key ($k_{E_aE_b}$) between $\id_a$
and $\id_b$. Similarly, \textit{B} computes the same shared key, completing the
connection establishment. This symmetric shared key is then used to encrypt
data packets between the two hosts.

\subsubsection{Encrypted Communication}

After the connection establishment, communication is based on symmetric
cryptographic operations.  First, the host uses the symmetric key that it
shares with the peer to encrypt the packets to the peer.  Any existing
CCA-secure encryption scheme can be used. Second, the host computes a MAC for
every packet that it sends, using the symmetric key that is shared with its AS
($k_{HA}$). This allows the host's AS to link every packet to its source and to
drop packets from (potentially) malicious hosts.

\subsubsection{Data Forwarding}
\label{sssec:data_forwarding}

Forwarding operations at \ers in source ASes ensure that only packets from
authenticated hosts and authorized \eids leave the source AS. Border routers in
destination ASes forward packets to the correct hosts based on the destination
\eids.  Transit ASes do not perform additional operations and simply forward
packets to the next AS on the path. As per our design choice, only symmetric
cryptographic operations are used, enabling a high-performance data forwarding.

Recall that communication end-points are specified as \AID:\eid tuples.  For
inter-domain forwarding, \ers use \AID to forward packets.  Specifically, for
external packets entering the AS, an \er checks whether the packet has arrived
at the destination AS. If not, the packet is forwarded to the neighboring AS
towards the destination AS. At the destination AS, the \er checks the following
conditions: 1)~the destination \eid ($\id_d$) has not expired, 2)~$\id_d$ has
not been revoked, and 3)~$HID_D$ is valid (\ie is registered and non-revoked).

If all conditions are satisfied, then the packet is forwarded to the
destination host: \ers derive the corresponding HID from the \eid and then
forward the packet; we assume that intra-domain routers forward packets based
on HIDs (\eg IP addresses).

\begin{figure}[!t]
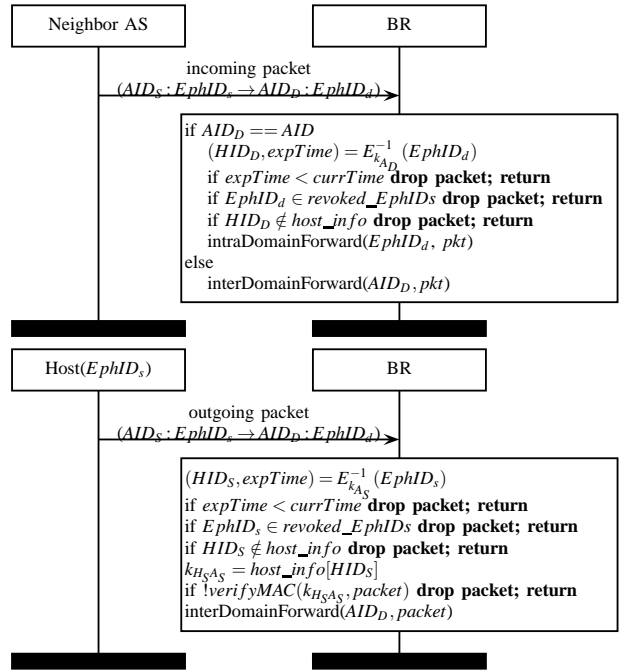

{\scriptsize
    \hspace*{0.4cm}
	\begin{minipage}[b]{1.1\linewidth}
	$\mathbf{\AID:}$ \AID of the Destination AS\\
    $\mathbf{\id_s, \AID_S:}$ Source \eid and \AID in the packet\\
    $\mathbf{\id_d, \AID_D:}$ Destination \eid and \AID in the packet\\
    $\mathbf{revoked\_ids:}$ List of revoked \eids\\
    $\mathbf{verifyMAC(k, M):}$ Verifies MAC of message $M$ using $k$\\
	\end{minipage}
	\hspace*{0.4cm}
	\begin{minipage}[b]{1\linewidth}
		\begin{msc}{}
		\drawframe{no}
		\setlength{\topheaddist}{0cm}
		\setlength{\bottomfootdist}{0cm}
		\setlength{\instdist}{4cm}
		\setlength{\envinstdist}{1cm}
		\setlength{\instwidth}{2.3cm}
		\declinst{in}{}{Neighbor AS}
		\declinst{er}{}{BR}
		\nextlevel[-2.5]
		\mess{\parbox{3.5cm}{\centering incoming packet\\($\AID_S:\eid_s \to \AID_D:\eid_d$)}}{in}{er}
		\nextlevel[0.5]
        \setlength{\actionwidth}{5.8cm}
        \setlength{\actionheight}{2.5cm}
		\action{
			\begin{minipage}{5.7cm}
			if {$\AID_D == \AID$}\\
			\hspace*{0.3cm}$(HID_D, expTime) = E^{-1}_{k_{A_D}}(\id_d)$\\
			\hspace*{0.3cm}if {$expTime < currTime$} \textbf{drop packet;} \textbf{return}\\
			\hspace*{0.3cm}if {$\id_d \in revoked\_{\id}s$} \textbf{drop packet;} \textbf{return}\\
			\hspace*{0.3cm}if {$HID_D \notin host\_info$} \textbf{drop packet;} \textbf{return}\\
			\hspace*{0.3cm}intraDomainForward($\id_d$, $pkt$)\\
			else\\
			\hspace*{0.3cm}interDomainForward($\AID_{D}, pkt$)
			\end{minipage}
		}{er}
		\nextlevel[4.7]
		\end{msc}
	\end{minipage}
	\hspace*{0.4cm}
	\begin{minipage}[b]{1\linewidth}
        \vspace{5pt}
		\begin{msc}{}
		\drawframe{no}
		\setlength{\topheaddist}{0cm}
		\setlength{\bottomfootdist}{0cm}
		\setlength{\instdist}{4cm}
		\setlength{\envinstdist}{1cm}
		\setlength{\instwidth}{2.3cm}
		\declinst{out}{}{Host\\($\id_s$)}
		\declinst{er}{}{BR}
		\nextlevel[-2.5]
		\mess{\parbox{3.5cm}{\centering outgoing packet\\($\AID_S:\eid_s \to \AID_D:\eid_d$)}}{out}{er}
		\nextlevel[0.5]
        \setlength{\actionwidth}{5.8cm}
        \setlength{\actionheight}{2.3cm}
		\action{
			\begin{minipage}{5.7cm}
			$(HID_S, expTime) = E^{-1}_{k_{A_S}}(\id_s)$\\
			if {$expTime < currTime$} \textbf{drop packet;} \textbf{return}\\
			if {$\id_s \in revoked\_{\id}s$} \textbf{drop packet;} \textbf{return}\\
			if $HID_S \notin host\_info$ \textbf{drop packet;} \textbf{return}\\
			$k_{H_SA_S}=host\_info[HID_S]$\\
			if {$!verifyMAC(k_{H_SA_S},packet)$} \textbf{drop packet;} \textbf{return}\\
			interDomainForward($\AID_{D}, packet$)
			\end{minipage}
		}{er}
		\nextlevel[4.4]
		\end{msc}
	\end{minipage}
}
\caption{Procedures for Data Packet Forwarding at \ERs for Incoming~(Top) and
Outgoing~(Bottom) Packets.}
\label{fig:forwarding}
\end{figure}

For outgoing packets, an \er forwards the packets to a neighboring AS only if
all of the following conditions are satisfied: 1)~the source \eid~($\id_s$) has
not expired, 2)~$\id_s$ has not been revoked, 3)~$HID_S$ is valid, and 4)~the
MAC in the packet is correct.

To verify the MAC in the packet, an \er retrieves the shared key~($k_{HA}$)
between the source host and the AS by searching the host information database
($host\_info$) using the $HID$ of the source host as the key. These checks
ensure that only authenticated packets leave the source AS.

\subsection{Shutoff Protocol}
\label{ssec:shut_off_protocol}

Shutoff protocols are designed to allow hosts to selectively block traffic from
specific source hosts. In our architecture, an \ADlow checks the validity of a
shutoff request and then blocks the source \eid. The agent checks whether a
customer-host has actually sent the specific packet that the requesting party
reports and whether the party is authorized to make the request (\ie the
requesting host was indeed the recipient of the specific packet). The agent
does not examine the intent of the source and whether the packet is malicious.

\begin{figure}[t]
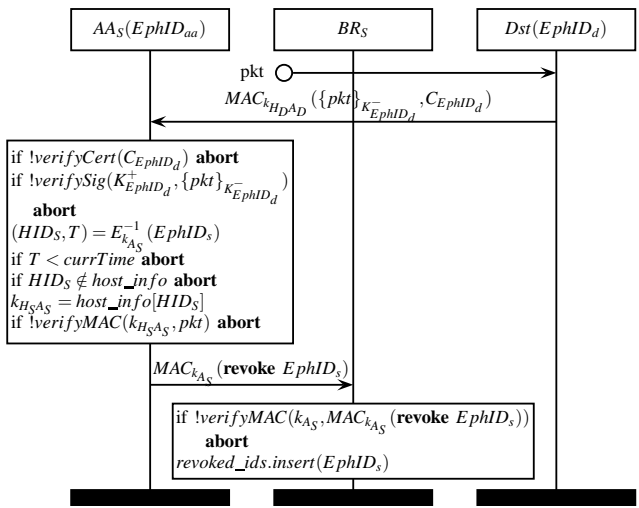

{\scriptsize
\hspace*{0.25cm}
\begin{minipage}[b]{1.1\linewidth}
$\mathbf{\id_s, \id_d}:$ Src/Dst \eids in the packet\\
$\mathbf{Dst}:$ Dst Host (\ie Host that is using $\id_d$)\\
$\mathbf{pkt}:$ packet that is sent by the Src Host but unwanted by the Dst Host\\
$\mathbf{\ADshort_S, BR_S}$: Accountability agent, \ER at Source AS\\
\end{minipage}
\hspace*{0.25cm}
\begin{minipage}[b]{1\linewidth}
\centering
\begin{msc}{}
\drawframe{no}
\setlength{\topheaddist}{0cm}
\setlength{\bottomfootdist}{0cm}
\setlength{\instdist}{2.7cm}
\setlength{\envinstdist}{1cm}
\setlength{\instwidth}{2.1cm}
\declinst{acc}{}{$\ADshort_S(\id_{aa})$}
\declinst{er}{}{$BR_S$}
\declinst{dst}{}{$Dst(\id_{d})$}
\nextlevel[-3.2]
\selfmesswidth=3.5cm
\found{}{pkt}{dst}[0.3]
\nextlevel[0.5]
\setlength{\actionwidth}{4.2cm}
\setlength{\actionheight}{0.7cm}
\nextlevel[0.8]
\mess{\parbox{3.4cm}{$MAC_{k_{H_DA_D}}(\{pkt\}_{K_{\id_d}^-},C_{\id_d})$}}{dst}{acc}
\nextlevel[0.5]
\setlength{\actionwidth}{3.8cm}
\setlength{\actionheight}{2.7cm}
\action{
    \begin{minipage}{3.7cm}
        \vspace*{0.2cm}
        if $!verifyCert(C_{\id_d})$ \textbf{abort}\\
        if $!verifySig(K_{\id_d}^+,\{pkt\}_{K_{\id_d}^-})$\\
        \hspace*{0.3cm}\textbf{abort}\\
        $(HID_S, T)=E^{-1}_{k_{A_S}}(\id_s)$\\
        if $T<currTime$ \textbf{abort}\\
        if $HID_S \notin host\_info$ \textbf{abort}\\
        $k_{H_SA_S}=host\_info[HID_S]$\\
        if $!verifyMAC(k_{H_SA_S},pkt)$ \textbf{abort}\\
    \end{minipage}
}{acc}
\nextlevel[6.5]
\mess{$MAC_{k_{A_S}}(\textbf{revoke }\id_s)$}{acc}{er}
\nextlevel[0.5]
\setlength{\actionwidth}{4.8cm}
\setlength{\actionheight}{1cm}
\action{
    \begin{minipage}{4.7cm}
        if $!verifyMAC(k_{A_S}, MAC_{k_{A_S}}(\textbf{revoke }\id_s))$\\
        \hspace*{0.3cm} \textbf{abort}\\
        $revoked\_ids.insert(\id_s)$
    \end{minipage}
}{er}
\nextlevel[1.5]
\end{msc}
\end{minipage}
}
\caption{Procedure for Shutoff Protocol.}
\label{fig:shut-off}
\end{figure}

Figure~\ref{fig:shut-off} shows the procedure for the shutoff request: the
destination host that owns $\id_d$ is attempting to block traffic coming from
$\id_s$ after receiving a specific packet.  The destination host creates a
shutoff request message with the following information: 1)~the received packet,
2)~a signature over the unwanted packet using the private key of $\id_d$
($K_{\id_d}^-$), and 3)~the certificate of $\id_d$. The packet is included as
evidence that the source has indeed sent traffic to the destination and the
shutoff request is not rogue; the signature and the certificate prove that the
destination host owns $\id_d$.  Then, the destination host sends the request
message to the \ADlow of the source host.

Upon receiving the request, the \ADlow verifies the certificate of $\id_d$ and
the signature in the request message to confirm that the request has indeed
been made by the destination host who owns $\id_d$.  Then, to ensure that the
packet has been actually generated by the source that owns $\id_s$, the agent
checks the authenticity of the packet using the shared key ($k_{H_SA_S}$) with
the source host. Finally, the \ADlow instructs the \ers to revoke $\id_s$ by
putting it into their $revoked\_ids$ list.

If misused, the shutoff protocol can be used to launch a DoS attack against a
benign source. To reduce the risk of such DoS attacks, we only authorize the
recipient of a packet to initiate a shutoff request (\ie the destination AS and
the destination host that owns the destination \eid). In
Section~\ref{sssec:strengthen_shut_off}, we discuss how other entities on the
communication path can be authorized to initiate a shutoff request.

\section{Implementation \& Performance Evaluation}
\label{sec:evaluation}
We present the implementation and performance evaluation of the core
architecture's components---the \eid management server and an \er.

\subsection{\eid Management Server}

The \eid Management Server (\ES) is responsible for generating \eids and for
assigning them to hosts. The \eid generation must be efficient since our
architecture should support even per-flow \eids. We describe the \eid
structure, the \ES implementation, and then evaluate the performance of the
\eid generation procedure.

\subsubsection{\eid Structure} \label{subsubsec:eid_struct}

\begin{figure}[h]
\centering
\includegraphics[width=.9\columnwidth]{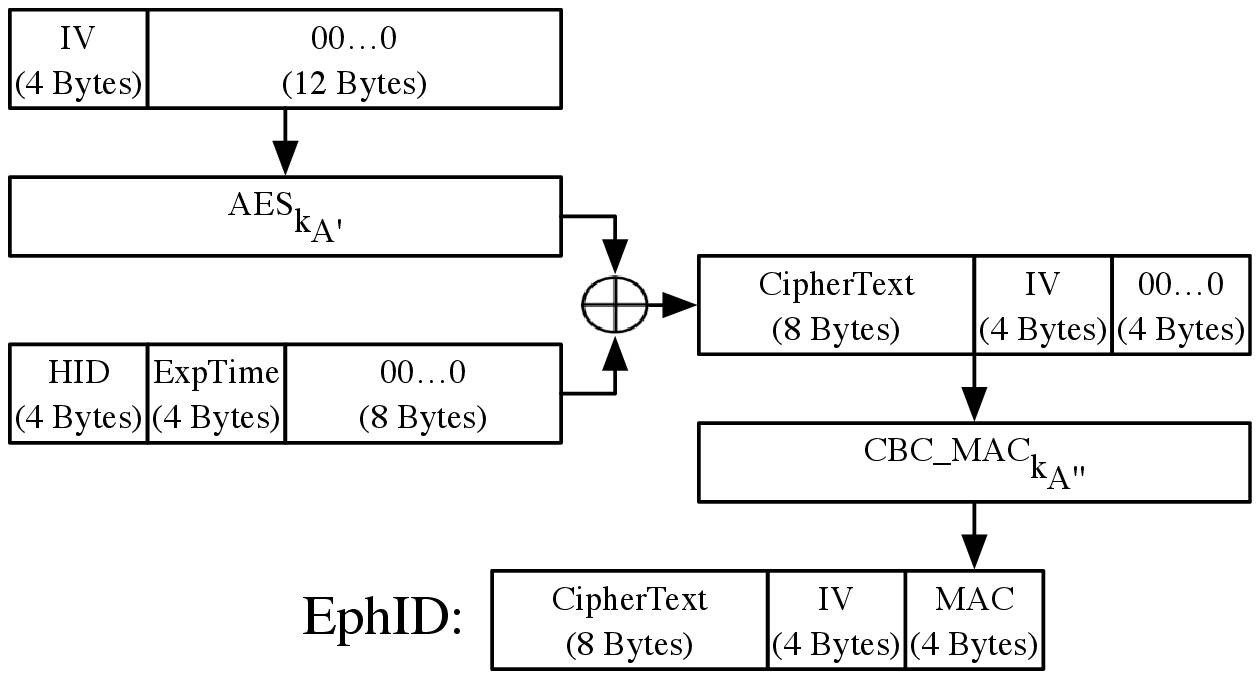}
\caption{\eid Construction.}
\label{fig:ephid}
\end{figure}

We engineer the \eid length to optimize for AES processing; AES operates on
16-byte~(B) blocks and is the only cipher with widespread hardware support.  

An \eid requires the HID of the host and an expiration time ($ExpTime$).  We
use 4\,B for the HID, which are sufficient to uniquely represent all hosts even
in large ASes.  The expiration time is 4\,B long, which allows us to use Unix
timestamps with one second granularity.  

Recall that the security requirement for \eids is a CCA-secure encryption
scheme.  To this end, we use a generic composition called
Encrypt-then-MAC~\cite{bellare2000-2} that combines a symmetric encryption with
a message authentication code (MAC) (Figure~\ref{fig:ephid}).  First, the
concatenation of $HID$ and $ExpTime$ is encrypted using AES in counter mode.
Secure operation of this mode requires a unique initialization vector (IV) for
every encryption (\ie for every \eid). Moreover, the use of the IV allows us to
generate multiple \eids for a single $HID$. Note that the plaintext data is
shorter than a single AES block (16\,B) and thus the input must be padded to
16\,B; the one-block plaintext requires a single AES operation.

Next, a message authentication tag is computed. The tag is computed over the
first 8\,B of the previously generated ciphertext and the IV that was used in
that encryption. We use CBC-MAC based on AES to generate the authentication
tag.

Finally, the \eid is constructed from the 8\,B of the ciphertext, 4\,B of the
IV, and 4\,B of the authentication tag (computed over the first two values);
the total length is 16\,B. Note that the keys used for encryption ($k_{A'}$)
and authentication ($k_{A''}$) are different; however, they can be derived from
the secret key of the AS ($k_A$).

\subsubsection{\ES Implementation} 

The \ES generates \eids according to the procedure in
Figure~\ref{fig:id-issue}.  For asymmetric cryptography, we use cryptographic
primitives based on Curve25519~\cite{bernstein2006}, which is proven to have
high performance and features small public-keys (32\,B) and small signatures
(64\,B). Key exchange is done using the elliptic-curve variant of
Diffie-Hellman (ECDH). To create digital signatures for certificates, we use
the ed25519 signature scheme~\cite{bernstein2012} and the ed25519 SUPERCOP
REF10 implementation\footnote{http://bench.cr.yp.to/supercop.html}.  For
symmetric cryptographic operations, we leverage Intel AES-NI~\cite{gueron2010}
-- a new encryption instruction set. Furthermore, we implement the host
database ($host\_info$) that stores the shared keys between hosts and the \AS
as a hashtable using HID as the key.

As an optimization, we parallelize the \eid generation by using 4 processes to
simultaneously handle \eid requests. The parallelization is straightforward since the
generation does not require any coordination (\eg shared memory or inter-process
communication) between the processes. However, no other optimizations were performed
(\eg optimizing the ed25519 REF10 implementation).

\subsubsection{\ES Performance Evaluation}
\label{sssec:RS_Performance}

We demonstrate the efficiency of generating per-flow \eids. To this end, we
need statistics for the peak flow generation rate inside an AS.

We use a 24-hour packet trace of HTTP(S) traffic from a major network provider
that manages network connections to universities and research facilities in an
European country.  The trace contains over 104 million and 74 million entries
for HTTP(S) traffic respectively. Each entry contains a timestamp and
anonymized source/destination IDs. We identify 1,266,598 unique hosts
generating a peak rate of 3,888 active HTTP(S) sessions per second.

We test our implementation on a desktop machine with an Intel Core i5-3470s CPU
(4 cores, 2.9GHz) and 4 GB of DDR3 memory. For 500,000 \eid requests, our
implementation runs for 6.9~seconds. On average, \SI{13.7}{\micro\second} are
needed for a single \eid generation, translating to a generation rate of 72.8k
\eids/sec --- over 18~times higher than the request rate. Our experiment shows
that even a low-end desktop machine can keep up with the traffic demands of a
real \AS that has over 1.2 million hosts.

\subsection{\ER}
\label{ssec:EdgeRouterImpl}

We describe our \er prototype starting with the structure of the network
header. Then, we describe the \er implementation and evaluate the forwarding
performance.

\subsubsection{\name Header Information}

\begin{figure}
\centering
\includegraphics[width=.9\columnwidth]{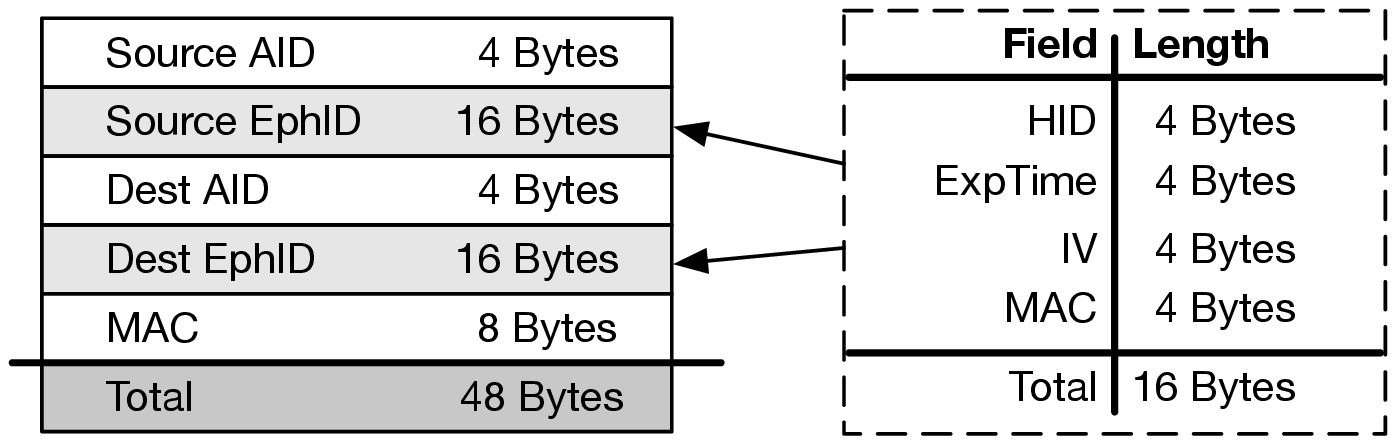}
\caption{\name Header Information and \eid Field Lengths.}
\label{fig:nh}
\end{figure}

The network header information (Figure~\ref{fig:nh}) contains the source and
destination end points (expressed as \AID:\eid tuples) and a MAC over the
packet's content. We use 4\,B to express the \AID since 4\,B are used for AS
numbers in the Internet; the \eid field requires 16\,B as described in
Section~\ref{subsubsec:eid_struct}; the MAC field requires 8\,B. The fields in
the packet header sum up to 48\,B.

\subsubsection{\ER Implementation}

Our \er performs additional processing compared to traditional IPv4/IPv6
forwarding (Figure~\ref{fig:forwarding}). Namely, the \er additionally performs
one decryption, two table lookups, and one MAC verification.

We use DPDK~\cite{dpdk} as our packet processing platform, which allows us to
implement the required functionality in userspace. The decryption of the \eid
in the packet is implemented through Intel AES-NI~\cite{gueron2010}.

\subsubsection{Forwarding Performance Evaluation}

We evaluate the forwarding performance on a commodity server with two Intel
Xeon E5-2680 CPUs and two non-uniform memory access (NUMA) nodes; each NUMA
node has four banks of 8~GB DDR3 RAM. The server is equipped with 6 dual-port
10~GbE NICs, providing a total capacity of 120~Gbps. To generate traffic, we
use Spirent-SPT-N4U-220~\cite{spirent} connected back-to-back with the server.
The server receives the traffic, processes it, and sends it back to the
generator.

\begin{figure}[!t]
    \centering
    \advance\leftskip-0.5cm
    \advance\rightskip-0.5cm
    \subfloat[][]{
        \includegraphics[width=0.47\columnwidth]{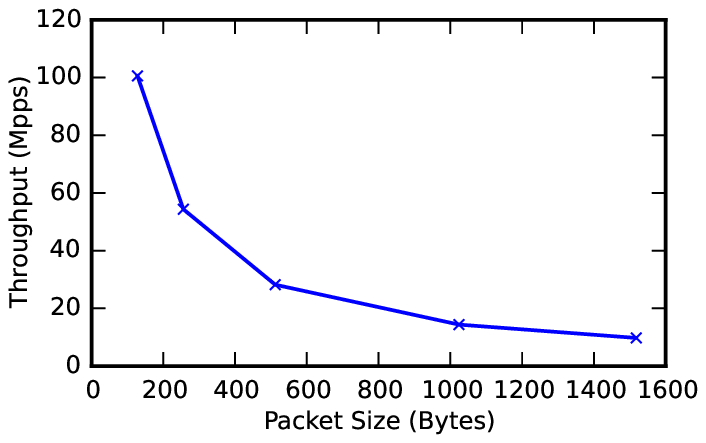}
        \label{fig:latency}
    }
    \subfloat[][]{
        \includegraphics[width=0.47\columnwidth]{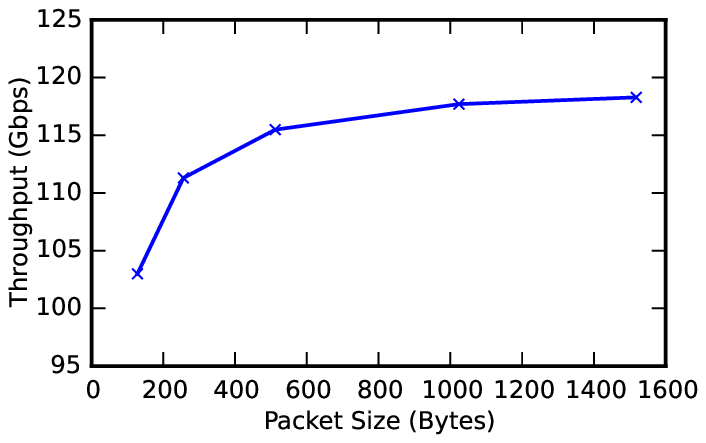}
        \label{fig:throughput}
    }

    \caption{Forwarding performance expressed as (a) packet-rate and (b)
    bit-rate.}
    \label{fig:fwd_perf}
\end{figure}

We perform a throughput experiment for 5 different packet sizes --- 128, 256,
512, 1024, and 1518-byte packets.  The results (Figure~\ref{fig:fwd_perf})
confirm that we are able to perform the required additional processing without
incurring a throughput penalty. The measured performance matches the
theoretical maximum performance; we omit this performance line for
demonstration purposes since the two lines match.  Figure~\ref{fig:fwd_perf}(a)
shows that even for small packet sizes (\ie high packet-rates), the \er
performs optimally. Figure~\ref{fig:fwd_perf}(b) shows that as packet sizes
increase, we saturate the capacity of 120~Gbps.  The \er performs optimally
because the additional operations are lightweight.  The \er's CPUs have
adequate capacity to perform this processing without degrading performance for
the given packet rates. Under higher packet rates, the heavier load would start
to degrade forwarding performance slightly.

\section{Security Analysis}
\label{sec:sec_analysis}
We demonstrate how \name prevents attacks that undermine source accountability
and data privacy.

\subsection{Attacking Source Accountability}

An adversary attacking source accountability
(Section~\ref{subsec:send_acc_def}) has three attack vectors at hand.

\paragraph{\eid Spoofing} The adversary can attempt to use an \eid that is
issued to another host (the spoofed victim). For instance, an adversary that
shares the same access port with the victim can sniff traffic and observe valid
\eids that are in use. However, using such an \eid is not sufficient since
every outgoing packet has to contain a MAC that is computed with the shared key
between the host and the host's AS. Without the corresponding shared key, the
adversary cannot create valid MACs, resulting in spoofed packets that are
dropped by the host's AS (additionally making the attack visible). Obtaining
the shared key requires compromising 	the host: the shared key is generated
with a DH key exchange between the host and the Registry Server
(Figure~\ref{fig:bootstrapping}), which means that the adversary needs the DH
private value that is used for the \eid generation; our adversary model does
not account for a compromised host.

An active adversary can attempt to obtain an \eid by pretending to be another
host.  However, such an attack is infeasible: the adversary not only needs to
learn the control \eid ($\id_h^{ctrl}$) of the victim, but also needs to learn
the shared key between the victim and the source AS.

\paragraph{Unauthorized \eid Generation} The adversary can attempt to create an
unauthorized \eid. However, such an attempt is not feasible since the \eid
construction (Figure~\ref{fig:ephid}) is CCA-secure.

We achieve a CCA-secure encryption scheme through two primitives. First, we use
symmetric encryption in counter mode with a fresh IV for every encryption; this
encryption is secure under a chosen plaintext attack. Second, we use a CBC-MAC
scheme to authenticate the concatenation of the ciphertext and the IV. Note
that our use of the CBC-MAC is secure against chosen plaintext attacks since
the input length to the CBC-MAC is fixed to 16\,B.\footnote{CBC-MAC is insecure
for variable-length messages~\cite{bellare2000-1}.}.  The combination of these
two primitives results in CCA-secure encryption scheme~\cite{bellare2000-2}.

\paragraph{Identity Minting} A common attack against systems that provide
accountability is identity minting, whereby a malicious host attempts to create
multiple (unauthorized) identities. In \name, since host identifiers (HIDs) are
generated by the AS and assigned only to the hosts that have authenticated, the
hosts cannot independently create multiple identifiers. In addition, if a host
requests a new HID, the previous HID and all associated \eids are revoked by
the \AS. Thus, at any moment every host on the network is identified by a
single HID.

\subsection{Attacking Privacy} \label{sssec:attack_on_privacy}

An adversary attacking data privacy can attempt to eavesdrop on communication
data or store it and decrypt it once he obtains the encryption keys. In \name,
traffic is encrypted by default and our scheme achieves perfect forward
secrecy: The symmetric key that is used for data encryption is bound to the
\eid (and the public/private key pair for that \eid) that is used for the
corresponding communication session.  This key pair is not used to derive other
encryption keys and is not derived from other long-term private keys
($K_{AS}^-$, $K_{H}^-$). Hence, only the compromise of a private key for an
\eid compromises data privacy and only for the communication session that uses
this \eid.

Alternatively, an AS-level adversary can actively try to compromise data
privacy of a customer host through a MitM attack. The malicious AS can perform
a MitM attack during the shared key establishment between the victim ($\eid_v$)
and a peer host ($\eid_p$).  In this attack the malicious AS replaces the
certificate for the \eid of the victim host ($C_{\eid_v}$) with another (fake)
certificate, pretending to be the victim host to the peer host; the peer host
accepts $C_{\eid_v}$. However, the AS cannot deceive the victim by pretending
to be the peer host because it cannot generate the certificate for $\eid_p$
($C_{\eid_p}$) that is signed by the private key of the peer host's AS.
Consequently, the connection is not established and the adversary cannot read
any communication of the hosts. The MitM attack is only possible if the source
and destination ASes collude, which we do not consider in our model.

For communication between two hosts in the same AS (\ie intra-domain
communication), \name does not provide any privacy guarantee from the AS: the
identities of the two hosts are already known to the AS (compromising host
privacy), and the AS can perform MitM attacks to decrypt communication between
the hosts (compromising data privacy) as the AS can fake both certificates for
the \eids that the hosts use.  The two hosts can use security protocols in
higher layers (\eg TLS) to encrypt the content of the communication.

\subsection{Other Attacks}

\paragraph{Unauthorized Shutoff Requests} The shutoff protocol can be misused
to perform a denial-of-service attack against a host. To prevent such an
attack, three measures are implemented to prevent unauthorized shutoff
requests. First, only the destination host and destination AS are authorized
to issue a shutoff request. Furthermore, the shut-off requester has to present
the unwanted packet that proves that the source has indeed sent the packet.
Since every packet has been cryptographically marked by the source AS, the
destination cannot make a shutoff request with a rogue packet. Lastly, the
shutoff requester must present its authorization credentials---it needs to sign
the request message with the private key associated with the destination \eid,
and include the corresponding short-lived certificate in the request message,
proving that it is an authorized party.

\paragraph{DDoS Attacks on Hosts} The architecture provides intrinsic defense
against DDoS attacks for three reasons. 1)~Since spoofing the source identifier
is difficult, reflection DDoS attacks (\eg DNS reflection) are difficult to
launch. 2)~The shutoff protocol allows the victim to suppress unwanted traffic.
3)~Due to strong accountability, the victim can ask the AS of malicious host to
take action against the host behind the \eids that are generating a lot of DDoS
attack traffic.

\section{Practical Considerations}
\label{sec:advanced}
\subsection{DNS Registration}
\label{ssec:dns_registration}

Today, the names of publicly accessible services (\eg an online shopping
website) are typically registered to public DNS servers. In \name, the servers
that host such services publish the \eid to a public DNS server, and the DNS
server returns the \eid with the corresponding certificate for a requested
domain name. To this end, the server performs two tasks: 1)~it requests an \eid
and the associated certificate from its \AS; and 2)~it registers the
certificate under the domain name to DNS;\footnote{We assume DNSSec to
authenticate DNS records.} the registered \eid will be used as the destination
address in future communication.

Publishing certificates to the DNS raises a problem: a shutoff request against
a published \eid would terminate any ongoing communication sessions that use
this \eid. A na\"{i}ve solution is to update the DNS entry with a new \eid
whenever the published \eid becomes invalid. However, this would become
burdensome for the DNS infrastructure if attackers continuously issue shutoff
requests against a domain.

Our solution is to define \textit{receive-only \eids}---\eids that are used
only to receive packets and are never used as the source \eids. Since they are
never used as the source identifier, they cannot become the target of shutoff
requests.  To avoid using receive-only \eids as the source identifier, the
communication establishment to a server needs to be changed (\ie the server
does not respond to the client using the receive-only \eid).

\paragraph{Client-Server Connection Establishment}~ To support receive-only
\eids by the server, the connection establishment procedure in
Section~\ref{ssec:conn_est} is extended. To simplify the narrative, assume that
the client uses $\id_c$ to connect to the server, and that the server uses
$\id_r$ as the receive-only \eid and $\id_s$ to serve the client.

After obtaining $\id_r$ from DNS, the client contacts the server using $\id_c$
and $\id_r$ as the source and destination \eids, respectively. The server
verifies the short-lived certificate of $\id_c$ and computes a shared key that
will be used to encrypt data packets between the client and the server.
However, instead of using the short-lived certificate of $\id_r$, the server
uses the short-lived certificate of $\id_s$ to compute the shared key. Then in
the response message to the client, the server includes the short-lived
certificate of $\id_s$ to inform the client that $\id_s$ will be used by the
server to serve the client.

The client verifies the short-lived certificate of $\id_s$ and computes the
shared key using the certificates for $\id_s$ and $\id_c$. In addition, the
client uses $EphID_s$ as the destination \eid to communicate to the server.

\paragraph{Protecting DNS Queries} Using the certificates for the two \eids
(\ie $\eid_c$ and $\eid_r$), DNS queries are encrypted just like any other data
communication. Hence, only the DNS server and the host knows the content of the
query (\eg domain name). However, if the DNS server is operated by the host's
\AS, the \AS can compromise the privacy of the DNS query---the AS knows the
identity of the host from the \eid and retrieves the content of the query from
the DNS server. To prevent such a compromise, the host can use a DNS server
that he trusts and that is not operated by the \AS that he resides in.

\paragraph{DNS Poisoning}
A malicious \AS can poison its local DNS servers with rogue entries. When the
victim attempts to connect to a certain domain, the \AS can successfully launch
a MitM attack.  We do not explicitly address DNS security since it is not a
network-layer issue. With \name, users can securely communicate
with a trusted DNS server of their choice, avoiding their \AS.

\subsection{Hosts Behind Connection-Sharing Devices}
\label{ssec:conn-sharing}

In the Internet, connection sharing devices (\eg NAT) are often used.  For
example, DSL or cable modems often have wireless Access Point functionality
that allows multiple devices (\eg laptops, smart phones) to connect to the
Internet; and, Internet caf\'es share their Internet connection and make it
accessible to their customers. In this section, we describe two approaches that
embrace connection-sharing devices in \name. For brevity, a connection sharing
devices is referred to as an Access Point (AP).

\paragraph{Bridge-mode} In this approach, the AP serves as a transparent bridge
that interconnects users behind the AP to the \AS. The \AS requires all users
to be directly authenticated to itself. In this approach, the \AS needs to
authenticate every single user, even those that may stay in the \AS network for
only a short period of time. Alternatively, the \AS can delegate the management
of connection sharing to the corresponding APs.

\paragraph{NAT-mode} In this approach, the AP creates a small domain of its own
while acting as a host to the \AS network. That is, the AP performs the
protocol described in Section~\ref{sec:details} as a host to the \AS while
playing the roles of a RS, an \ES, a router, and an \ADlow on behalf of its
clients.

As a \RS, the AP bootstraps the hosts into the AP's internal network: it
authenticates the hosts to the internal network, negotiates shared keys that
are used to authenticate the packets that the hosts send, and provides
bootstrapping information.

As an \ES, the AP makes \eid requests on behalf of its hosts to the \AS. The
procedure that the AP follows to acquire \eids for its hosts is similar to the
\eid issuance protocol described in Figure~\ref{fig:id-issue}, but with two
differences.  First, when requesting for an \eid to the \ES of the \AS, the AP
uses an ephemeral public key that is supplied by its host. Second, the AP keeps
track of the \eids that are assigned to the hosts as a list, \ie $\eid\_info$
(as opposed to deriving HIDs from \eids) since \eids are encrypted using the
\AS's secret key and \eids contain HIDs assigned to the AP, not to its hosts.
This list is used to identify the hosts using \eids in the packets.

As a router, the AP implements the data forwarding procedures described in
Figure~\ref{fig:forwarding}, but with two differences. First, instead of
parsing the \eids to determine the HID of the host, the AP uses the $\id\_info$
list. Second, for outgoing packets, in addition to verifying the MAC in the
packets using the shared keys with its hosts, the AP replaces the MAC using its
shared key with the AS before forwarding the packets to the AS.

Finally, as an \ADlow, the AP identifies the misbehaving hosts based on \eids.
Since the hosts behind an AP are not visible to the \AS and since the \AS
issues \eids to the AP not to the hosts, the \AS holds the AP accountable for
misbehaving \eids. Then, the AP determines the host that is using the
misbehaving \eid.

\subsection{Connection Establishment Latency}

In Section~\ref{ssec:conn_est}, we described how two hosts establish a
connection with each other: using the short-lived certificates for the two
\eids, the two hosts compute the symmetric shared key that is used for data
encryption. The connection establishment requires one Round Trip Time (RTT)
before any communication can take place; however, this RTT can be eliminated.
On the very first packet of the connection establishment, the host encrypts its
data after computing the shared key. Then, the receiving host decrypts the data
after computing the shared key.

In the case of client-server communication, the connection establishment
(described in Section~\ref{ssec:dns_registration}) requires 1.5 RTTs, but this
latency can be reduced to 0 or 0.5 RTT depending on the desired level of data
privacy. If the client encrypts data on the first packet using the shared key
between her \eid and the receive-only \eid of the server, the client incurs
zero RTT in connection establishment. However, if an adversary compromises the
private key of the receive-only \eid, the adversary can decrypt the first
packets of all communication sessions with the server. The compromise of the
first packets is eliminated if a client does not send data on the first packet;
however, this incurs latency penalty of 0.5 RTT.

\subsection{Deployment in the Internet}

Although our ideas are not restricted to a certain Internet architecture, \name
can be used in today's Internet. To this end, the Generic Routing Encapsulation
(GRE) protocol~\cite{rfc2784}, which uses the IPv4 network as virtual
Point-To-Point (PTP) links to encapsulate various network protocols (\eg
tunneling IPv6 sites over IPv4 network) can be used. In \name, GRE
encapsulation can interconnect two \name entities, \eg between two \name
routers over IPv4 network.

\begin{figure}[h]
\centering
\includegraphics[width=.9\columnwidth]{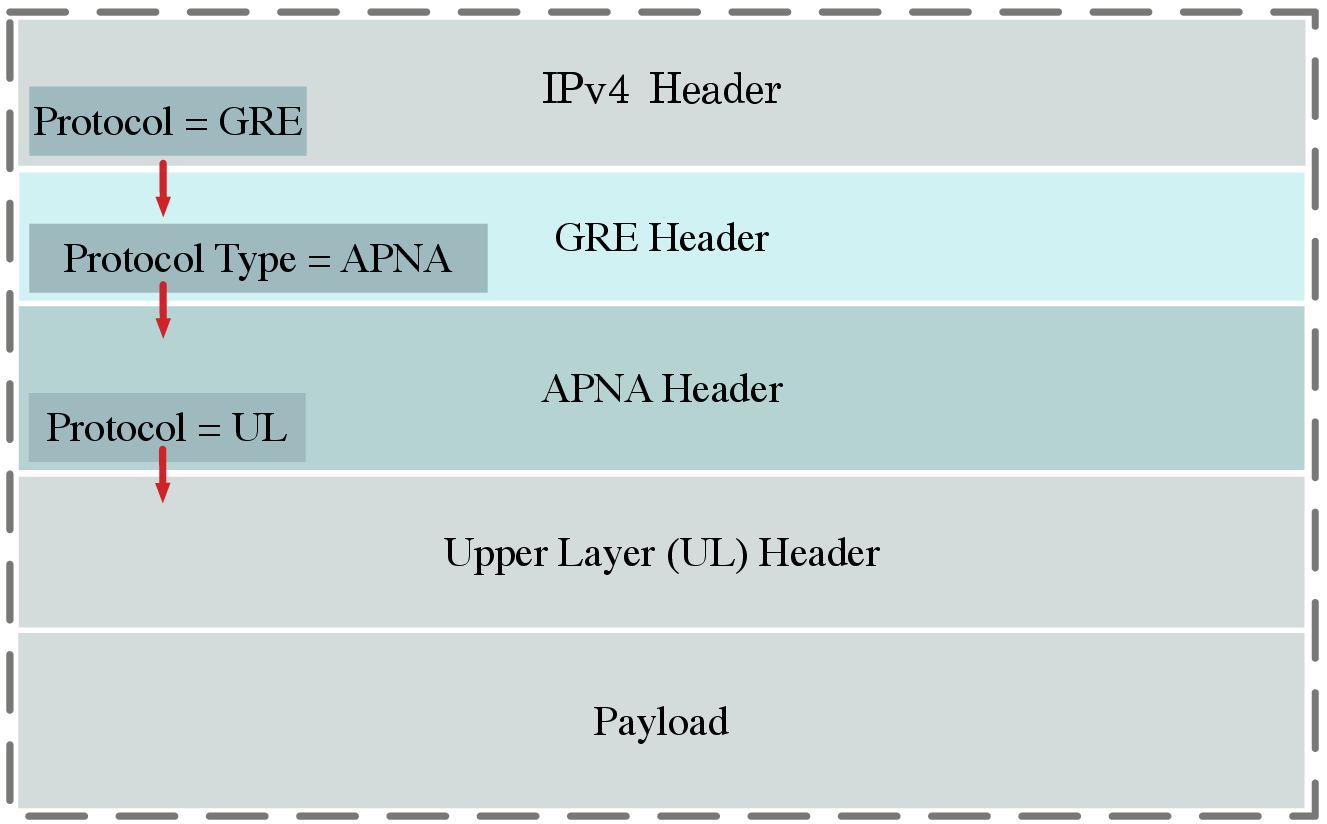}
\caption{\name Packet Structure.}
\label{fig:pkt_structure}
\end{figure}

Figure~\ref{fig:pkt_structure} shows the \name packet structure using the GRE
protocols. The source and destination addresses of the IPv4 header are that of
the two \name entities that are sending and receiving the \name packet. The
\name header and the data payload follow the GRE tunnel header; the GRE tunnel
specifies the encapsulated network protocol using the \textit{Protocol Type}
field. Since the GRE protocol uses the \textit{EtherType} numbers for
identifying the encapsulated protocol, we would need to request a dedicated
EtherType number from IANA.

IPv4 addresseses of the hosts serve as the HIDs in the IPv4 deployment. Note
that IPv4 addresses and HIDs are both four bytes long, and IPv4 addresses are
uniquely assigned to hosts in an \AS. In addition, IPv4 addresses of \name
routers serve as AIDs so that inter-domain routing continues to be based on
IPv4 addresses.

For intra-domain forwarding in the source AS, the source host puts its IP
address and the IP address of an \name router as the source and destination
addresses in the IPv4 header (that comes before the GRE header), respectively.
For intra-domain forwarding in the destination AS, an \name router 1)~decrypts
the destination \eid in the \name header to get the HID (\ie IPv4 address) of
the destination host; and, 2)~replaces the destination IPv4 address of the IPv4
header with the HID.

This intra-domain forwarding has a privacy implication. Within the source and
destination ASes, the addresses of the hosts are visible; hence, it is not
possible to provide any privacy guarantee against an adversary who observes
packets within the ASes. However, once an AS fully deploys \name (\ie all
routers forward packets based on \eids), this privacy implication disappears.

For inter-domain forwarding in the source AS, a \name router replaces
the addresses in the IPv4 header of the \name packet with its IPv4 address and
the destination \AID as the new source and destination addresses, respectively.
For all transit ASes, the packet is forwarded based on the destination address in
the IPv4 header.

\paragraph{\name Gateway} Making modifications to the host network stack is an
onerous task that hampers deployment of novel architectures. Hence, we propose
using \name gateways to bridge between the Internet and \name without having to
change the host network stack. An \name gateway has two roles: 1)~as
an \name host, it runs the protocols described in Section~\ref{sec:details};
and 2)~as a packet translator, it converts between native IPv4 and
\name packets. Assuming that the gateway uses different source \eid for
different IPv4 flows, the challenge in translating between IPv4 and \name
packets is determining the mapping between IPv4 flow information (identified by
the standard 5-tuple) in IPv4 packets and \name flow information (identified by
source and destination \AID:\eid pair) in \name packets.

When forwarding an outgoing IPv4 packet from a host to an \name router, the
gateway converts the IPv4 packet to a \name packet. IPv4 addresses in the \name
packet can be easily determined: the addresses of the gateway and the \name
router are used as the source and destination IP addresses, respectively.  In
addition, the source \AID:\eid information in the \name header can be easily
determined: for each new IPv4 flow, the gateway uses a different \eid.
However, determining the destination \AID:\eid is not trivial. In fact, the
gateway cannot determine the destination \AID:\eid solely based on the 5-tuple
information in the IPv4 packet from the host. 

Instead, the gateway has to rely on mechanisms that the host uses to find its
peer host. For instance, a client may use DNS that stores the short-lived
certificate and the IPv4 address of the server. The gateway that serves the
client learns the IPv4 address and the \AID:\eid of the server by inspecting
the DNS reply to the client. Then, the gateway uses the destination IPv4
address in the IPv4 packets from the client to the server to get \AID:\eid of
the server.

Note that the gateway can automatically learn the IPv4 address to \AID:\eid
mapping only if the host uses a well-known mechanisms (\eg DNS). Otherwise, the
host needs to statically configure the mapping between peer's IPv4 address and
the \AID:\eid pair into the gateway.

In the above client-server communication example, one may argue that the host
privacy of the server is lost since its IPv4 address is registered in DNS.  To
overcome such privacy loss, the IPv4 address can be removed from the DNS
record. When the client's gateway sees the DNS reply, it generates and appends
a random IPv4 address into the DNS reply. Then, based on the destination IPv4
address in the client's packets to the server, the gateway determines the
\AID:\eid of the server.


When forwarding an incoming \name packet from an \name router to a host, the
gateway needs to convert it to an IPv4 packet by choosing appropriate source
and destination IPv4 addresses. If the gateway already has the mapping between
the \name flow tuple and the IPv4 flow tuple (\ie the receiving host has sent
an outgoing packet with the IPv4 flow tuple), the gateway uses the IPv4 flow
tuple to create the IPv4 packet. If the gateway does not have the mapping, the
gateway needs to carefully choose the source and destination IPv4 addresses for
the IPv4 packet.

When choosing the source address, the gateway needs to ensure that the host can
distinguish between different flows. That is, every \name flow tuple must be
mapped to a unique IPv4 flow tuple. If the gateway uses its IPv4 address as the
source address in the IPv4 packet, two different \name flows may have the same
5-tuple information when they use the same source port number. Alternatively,
we define a \textit{virtual end-point} which consists of an IPv4 address (\eg
randomly drawn from a private address space), and the source port number in the
transport header in the \name packet.  The gateway assigns unique virtual
end-point for each \name flow, and the IPv4 address of the virtual end-point is
used as the source IPv4 address in the IPv4 packet.

To determine the destination IPv4 address, the gateway uses the destination
\eid information in the \name header. However, the mapping between \eid and
IPv4 address exists only if the destination has sent an outgoing packet or the
destination host has registered the mapping between its \eid and IPv4 address.
For example, a server administrator registers a (receive-only \eid, IP
address)-tuple to his gateway after registering his domain information in DNS.

\section{Discussion}
\label{sec:discussion}
\subsection{Ephemeral ID Granularity} \label{subsec:eid_gran}

Thus far, we have argued that \name does not impose the granularity at which
\eids should be used and we have shown that the \eids can be generated at high
speed (See Section~\ref{sssec:RS_Performance}). In this section, we present
four granularities at which \eids can be used.

\paragraph{Per-Flow Ephemeral ID} This is the typical use case where a host
uses different \eids for different flows. There are two advantages to per-flow
\eids. It prevents an observer's attempt to identify a common sender of
multiple flows by inspecting the content of the packets (\ie \name header and
payload). Shut-off incidents have limited impact on a host. It terminates the
flow that uses the reported \eid as the source; however, all other flows remain
intact. The disadvantage of this case is that a host needs to acquire and
manage \eids for every new flow.

\paragraph{Per-Host Ephemeral ID} On one end of the spectrum, a host
uses a single \eid for all packets. The advantage of this model is that a host
only needs to acquire and manage one \eid. However, there are two drawbacks.
Since all packets have the same source \eid, all packets are linked to a common
sender.  Shut-off incident terminate all connections from the host.

\textbf{Per-Packet Ephemeral ID:} A host could use different \eids per each
packet. Hence, it would be difficult to link different packets even to a single
flow, providing the strongest privacy guarantee. However, even the destination
host cannot demultiplex packets into flows based on the \name headers in the
packets. An additional protocol is necessary to demultiplex
packets~\cite{lee2016}.

\paragraph{Per-Application Ephemeral ID} An \eid can be used to represent all
packets that are generated by an application or a service that is running on
the host.  This \eid granularity facilitates managing traffic that are
generated by an application. For example, if an AS enforces its hosts to use
per-application \eids, the AS and its hosts could collaboratively identify
malicious applications (\eg DDoS bot application) that are running at the
hosts. The network identifies malicious activities (\eg creating flooding
attacks) to a source \eid and inform the host about the \eid; then the host
identifies the application that uses the \eid and takes appropriate actions.

\subsection{Support for ICMP}
\label{ssec:icmp}

In \name, ICMP is available by default in most cases because the source host
can be reached using the source \eid that is present in host's
packets\footnote{If the source \eid expires immediately after the packet leaves
the source AS, the source \eid becomes invalid. However, we expect such case to
occur infrequently.}. Hence, using the source \eid in a packet, one can send an
ICMP message to the source host.

Sending an ICMP message follows the same procedure as sending a data packet to
another host. An entity (\eg router or host) that wishes to send an ICMP
message uses one of its \eid as the source address in the packet and computes
the MAC using the shared key with its AS. \name offers privacy and
accountability for the ICMP messages: identity of the entity remains hidden
except to its AS, and the packet is authenticated by the AS. Consequently, if
an ICMP message is deemed to be faulty by the receiving host, he can hold the
ICMP message sender accountable for the message via the AS of the sender.

Unlike data communication between two hosts, however, the payload of ICMP
messages are not encrypted. Encrypting the payload is difficult because the
ICMP message sender cannot easily obtain the short-lived certificate of the
source \eid in the original message that have prompted the ICMP message. One
naive approach is to store short-lived certificates of all flows that the (ICMP
message) sender sees; however, this approach incurs a lot of storage overhead
to the sender. As our future work, we are exploring ways to encrypt ICMP
messages without imposing excessive overhead to the ICMP message sender.

\subsection{Strengthening the Shutoff Protocol}
\label{sssec:strengthen_shut_off}

If designed incorrectly, the shutoff protocol can be abused as a tool to
perform DoS attacks against benign hosts. Thus, it is important to correctly
identify the entities that are authorized to perform a shut-off.

In Section~\ref{ssec:shut_off_protocol}, we restricted the authorized parties
as the destination host and AS since these are the only two parties that will
provably receive the packet based on the \name header. However, there are
proposals to encode the forwarding paths into the packets (\eg Packet
Passport~\cite{liu2008}, ICING~\cite{naous2011}, and OPT~\cite{kim2014}). When
such proposals are combined with our architecture, the list of authorized
entities can be extended to include on path-ASes (or their routers),
strengthening the shut-off protocol.

\subsection{Handling Replay Attacks}

A malicious entity that aims to ``harm'' a source host may replay
packets of the source. In the short-term, replayed packets may induce shutoff
incidents against the source host, disrupting communication of the source; and
in the long-term, the AS of the source host may take retributive action against the
source host for repeated shutoff incidents.

Replay attacks can be prevented by making every packet unique. That is, a nonce
field is added to the \name header (Figure~\ref{fig:nh}), and a source host
puts a unique number for each generated packet. Then, the destination host
performs replay detection based on the nonces in the packets and discards  all
duplicate packets.

Ideally replayed packets should be filtered near replay location, but
this requires routers in the network to perform replay detection. Designing a practical
in-network replay detection mechanism that does not affect routers' forwarding
performance is not trivial; it is our future work to design such a
mechanism.

\subsection{\name-as-a-Service}
\label{ssec:name_as_a_service}

An ISP can offer \name{}'s accountability and privacy protection not only to
hosts in its network, but also to its downstream (\eg customer) ASes. In this
deployment, a downstream AS can be viewed as a connection-sharing device that
provides \name connections to its hosts. Then the downstream AS can work as a
\textit{transparent bridge} or \textit{NAT} to connect its customers to the ISP
(See Section~\ref{ssec:conn-sharing} for details).

\name-as-a-Service offers benefits to both the ISP and the downstream ASes. The
ISP can expand its \name customer base beyond its network. However, note that
the ISP can only offer \name-as-a-Service to ASes whose packets must go through
the ISP. This restriction is necessary since the ISP needs to be able to verify
all packets that are originating from the downstream ASes to act as the \ADlow.
The customer ASes, especially the small ASes that do not have a large number of
hosts (\ie small anonymity set), can enjoy stronger level of host privacy
protection by mixing with customers of other (upstream) ISPs.

However, there are challenges in deploying \name-as-a-Service. For example,
authentication process of the end-hosts become more complicated since the hosts
of the downstream AS may need to authenticate remotely. In addition, routing in
the downstream ASes become complex, especially for the ASes that are
multi-homed (e.g., managing \eids). As our future work, we are investigating
the challenges associated with offering \name-as-a-Service.

\subsection{Interaction with TLS}

\name by design addresses network layer security issues: (1) it prevents source
spoofing by imposing strict authentication of packets; and (2) it provides
communication privacy by hiding the identities of communicating parties and
supporting pervasive end-to-end encryption. However, \name does not deal with
security issues at higher layers (e.g., authenticating domain ownership).

\name can work in conjunction with security protocols that deal with security
issues at higher layers. For example, TLS can be implemented on top of the
encrypted end-to-end path between two hosts to perform user authentication.
However, not all functionalities of upper layer security protocol may be
necessary. For instance, since \name already provides a secure end-to-end
channel between hosts, the mechanism to establish a symmetric shared key for
data encryption may be omitted when implementing TLS on top of \name.

\subsection{Parameter Considerations}

\subsubsection{Expiration Time for \eids}

There are multiple factors to consider when deciding the expiration time for
\eids and the associated short-lived certificates: it should be sufficiently
long so that an \eid does not expire before the communication that uses the
\eid terminates. At the same time, it should be kept short so that \eid does
not last long beyond the end of the communication.

If \eids are used per flow, the expiration time can be set to 15 minutes as
98\% of the flows in the Internet last less than 15
minutes~\cite{Brownlee2002}. Alternatively, the \eid Issuance protocol (Section
\ref{ssec:ephID_issue}) can be extended to allow hosts to express their choice
of expiration time. For instance, an AS may specify three categories
(short-term, medium-term, long-term \eids) to accommodate diverse nature of
flow duration time.

\subsubsection{Managing Revoked \eids}
\label{sssec:revoked_eid}

\eids can be preemptively revoked before they expire: a host could revoke an
\eid that is no longer needed, or an \eid could have been subjected to a
shutoff incident. Regardless of the reason for revoking \eids, \ers in the ASes
need to store a list of revoked \eids (\ie $revoked\_\eids$ in
Figure~\ref{fig:forwarding}). If there are too many revocations in an AS, it
burdens the \ers since the size of the $revoked\_\eids$ would become large.

There are two ways to manage the size of $revoked\_\eids$ list. First, since
\eids will expire over time and packets using expired \eids are dropped, the
expired \eids can be removed from $revoked\_\eids$. Second, if too many \eids
of a host are revoked, \AS should view it as a sign of malicious activity by
the host. In such event, \AS revokes the HID of the host invalidating all \eids
that are issued to the host, and \AS assigns a new HID to the host. In
addition, the \AS can contact the host for corrective measures.

Such measures against malicious hosts by the \ASs are not radical.  Already in
today's Internet, ISPs that participate in Copyright Alert System
(CAS)~\cite{cas} actively take actions against the customers who repetitively
upload copyrighted contents illegally: a customer receives warnings up to 6
reported incidents of illegal uploads, and on the 7th incident, his ISP take
actions against the customer (\eg temporarily reduce connection bandwidth, take
educational course about copyright laws). In \name, an \AS can set a maximum
number of \eids that can be preemptively revoked for each host. Then if a host
exceeds the maximum number, the \AS can take actions against the host.

\subsection{Governments and Communication Privacy}

Although generally perceived as a threat on communication privacy, there are
legitimate reasons for governments to subvert communication privacy (\eg to
monitor terrorist activities). In fact, many governments by law mandate ISPs to
keep record of their Internet traffic (\eg source and destination IP addresses,
payload of the packets, etc).

\name protects communication privacy making mass surveillance difficult;
however, at the same time, it allows entities, such as a government, to
deanonymize communication when necessary. With the cooperation of an \AS, a
government can deanonymize the identity of hosts from \eids.  Furthermore, if
the government has cooperation from the \ASs in which communicating hosts
reside, the \AS could decrypt ongoing communication by performing a MitM
attack. However, the government cannot simply collect packets in the Internet
to observe communication since packets are encrypted (\ie making mass
surveillance difficult). In addition, since \name achieves perfect forward
secrecy, governments cannot decrypt all communication of a host, even if after
compromising the long-term public key of the host.

\section{Related Work}
Persona~\cite{mallios2009} is the first proposal to introduce the idea of
balancing privacy and accountability at the network layer. The source ISP
replaces the IP address of each outgoing packet with another address from an
assigned pool. Although this approach hides the source's identity, it breaks
the notion of flow and prevents the destination from demultiplexing
connections.

Accountable and Private Internet Protocol (APIP)~\cite{naylor2014} proposes an
architecture that balances accountability and privacy at the network layer. In
APIP, the source address in the network header is replaced with the address of
an \emph{accountability delegate} that vouches for the source's packets. The
return address can then be specified at a higher layer -- invisible from the
network --  protecting the source's privacy. Senders are expected to brief each
packet to their accountability delegate such that on-path devices can request a
``vouching proof'' from the corresponding delegate. 

APIP balances privacy and accountability at the network layer, but it comes
with certain limitations. APIP's notion of privacy is limited to sender-flow
unlinkability, leaving data privacy and the associated challenges (\eg key
distribution, management, and establishment) unaddressed. Our proposal presents
a holistic architecture that addresses these constraints and by default
supports data privacy.  Furthermore, the design of APIP precludes every packet
from being accounted for in the network: it is possible for a malicious host to
omit reporting packets to its accountability delegate when the flow for those
packets has been ``whitelisted''.\footnote{Verifiers do not verify flows that
have been ``whitelisted,'' and a sender does not brief packets unless it is
asked by its accountability delegate under the recursive verification method
(Section 5 in APIP~\cite{naylor2014}).} In \name, every packet is linked to its
sender since a MAC is computed using the shared key between the AS and the host
for every packet (Section~\ref{ssec:data_comm}). Second, masking the return
address complicates getting messages from the network back to the source---the
messages must be redirected through the accountability delegate of the source;
the complexity of this functionality remains unaddressed.  \name allows the
network to send messages directly to the source while preserving host privacy
and the accountability properties~(Section~\ref{ssec:icmp}).

\textbf{Source Accountability:} The Accountable Internet Protocol
(AIP)~\cite{andersen2008} treats source accountability as a central
architectural principle. In AIP, self-certifying IDs and a shutoff protocol
(implemented by smart Network Interface Cards) are used to identify and block
malicious sources. Our architecture uses self-certifying IDs in an
anonymity-preserving way and delegates the shutoff functionality to the source
domain.

In Passport~\cite{liu2008}, OPT~\cite{kim2014} and ICING~\cite{naous2011},
Message Authentication Codes are used for each AS on the end-to-end path,
allowing on-path ASes to verify the authenticity of packets.

Bender \etal\cite{bender2007} were first to introduce the concept of \ADslow in
their Accountability-as-a-Service (AaaS) proposal. However, AaaS does not
address privacy considerations and requires symmetric keys between all AS
pairs.  

\textbf{Host Privacy and Anonymity:} Raghavan \etal\cite{raghavan2009} propose
ISP-wide NATs to hide the hosts' identities from entities in other ASes. We
borrow their motivation that the size of today's large ISPs provides
sufficiently large anonymity sets.  Onion routing~\cite{reed1998} and Mix
Networks~\cite{chaum1981} by design provide source anonymity at the cost of
source accountability. 

Han~\etal\cite{han2013} propose a cross-layer design that uses psuedonyms to
hide the user's identity. Similar to \name, the proposal allows the user to
choose the level of anonymity and it uses encryption to mask the identity of
the host in network addresses. However, the proposal falls short of being a
complete architecture that balances between accountability and privacy: it does
not consider pervasive data encryption and the associated challenges, such as
certificate management, and does not consider source accountability.

\textbf{Data Privacy:} Farrell and Tschofenig~\cite{rfc7258} argue that
pervasive monitoring  -- defined as the widespread and often covert
surveillance through intrusive gathering of communication information -- is a
widespread attack on privacy. In response, Kent~\cite{kent2014} proposes
pervasive encryption as a countermeasure against pervasive monitoring. In a
related effort, the Let's Encrypt\footnote{https://letsencrypt.org}
organization encourages the use of encrypted web traffic by issuing free TLS
certificates for web servers. Our proposal does not replace transport-layer
encryption, but rather promotes pervasive encryption to a fundamental design
tenet of the network layer. In addition, we propose a concrete solution for key
distribution, establishment, and management.

MinimaLT~\cite{petullo2013} proposes an architecture that supports pervasive
data encryption and achieves PFS at low latency; however, MinimalLT does not
consider source accountability. In Section~\ref{ssec:conn_est}, we show how our
architecture supports data privacy with PFS while enforcing source
accountability.

\section{Conclusions}
We propose \name, an architecture that resolves the accountability-privacy
tussle by enlisting ISPs as \ADslow and privacy brokers. As \ADslow, ISPs
authenticate hosts and their packets into the network; and as privacy brokers,
ISPs anonymize the identities of communicating parties and assist in the
establishment of shared keys for end-to-end data encryption.

By facilitating (and by enabling by default) pervasive encryption between
endpoints, \name can help frustrate adversaries conducting indiscriminate mass
surveillance. At the same time, \name can assist in lawful, targeted request
for subscriber communications, since ISPs can comply with data retention laws
by storing customer to \eid bindings as well as the packets. However, abuse of
such requests for information are minimized due to the perfect forward secrecy
of our scheme: even if host's public key is compromised, the secrecy and
integrity of previous communications remain untouched.

\bibliographystyle{IEEEtranS}
\bibliography{0-string,apna,rfc}

\end{document}